\documentstyle[twoside,epsfig]{article}

\catcode`\@=11
\long\def\@makefntext#1{
\protect\noindent \hbox to 3.2pt {\hskip-.9pt
$^{{\eightrm\@thefnmark}}$\hfil}#1\hfill}       

\def\thefootnote{\fnsymbol{footnote}}
\def\@makefnmark{\hbox to 0pt{$^{\@thefnmark}$\hss}}    

\def\ps@myheadings{\let\@mkboth\@gobbletwo
\def\@oddhead{\hbox{}
\rightmark\hfil\eightrm\thepage}
\def\@oddfoot{}\def\@evenhead{\eightrm\thepage\hfil
\leftmark\hbox{}}\def\@evenfoot{}
\def\sectionmark##1{}\def\subsectionmark##1{}}

\oddsidemargin=\evensidemargin
\renewcommand{\thefootnote}{\fnsymbol{footnote}}

\newcounter{sectionc}\newcounter{subsectionc}\newcounter{subsubsectionc}
\renewcommand{\section}[1] {\vspace{12pt}\addtocounter{sectionc}{1}
\setcounter{subsectionc}{0}\setcounter{subsubsectionc}{0}\noindent
    {\tenbf\thesectionc. #1}\par\vspace{5pt}}
\renewcommand{\subsection}[1] {\vspace{12pt}\addtocounter{subsectionc}{1}
    \setcounter{subsubsectionc}{0}\noindent
    {\bf\thesectionc.\thesubsectionc. {\kern1pt \bfit #1}}\par\vspace{5pt}}
\renewcommand{\subsubsection}[1] {\vspace{12pt}\addtocounter{subsubsectionc}{1}
    \noindent{\tenrm\thesectionc.\thesubsectionc.\thesubsubsectionc.
    {\kern1pt \tenit #1}}\par\vspace{5pt}}
\newcommand{\nonumsection}[1] {\vspace{12pt}\noindent{\tenbf #1}
    \par\vspace{5pt}}

\newcounter{appendixc}
\newcounter{subappendixc}[appendixc]
\newcounter{subsubappendixc}[subappendixc]
\renewcommand{\thesubappendixc}{\Alph{appendixc}.\arabic{subappendixc}}
\renewcommand{\thesubsubappendixc}
    {\Alph{appendixc}.\arabic{subappendixc}.\arabic{subsubappendixc}}

\renewcommand{\appendix}[1] {\vspace{12pt}
        \refstepcounter{appendixc}
        \setcounter{figure}{0}
        \setcounter{table}{0}
        \setcounter{lemma}{0}
        \setcounter{theorem}{0}
        \setcounter{corollary}{0}
        \setcounter{definition}{0}
        \setcounter{equation}{0}
        \renewcommand{\thefigure}{\Alph{appendixc}.\arabic{figure}}
        \renewcommand{\thetable}{\Alph{appendixc}.\arabic{table}}
        \renewcommand{\theappendixc}{\Alph{appendixc}}
        \renewcommand{\thelemma}{\Alph{appendixc}.\arabic{lemma}}
        \renewcommand{\thetheorem}{\Alph{appendixc}.\arabic{theorem}}
        \renewcommand{\thedefinition}{\Alph{appendixc}.\arabic{definition}}
        \renewcommand{\thecorollary}{\Alph{appendixc}.\arabic{corollary}}
        \renewcommand{\theequation}{\Alph{appendixc}.\arabic{equation}}
        \noindent{\tenbf Appendix \theappendixc #1}\par\vspace{5pt}}
\newcommand{\subappendix}[1] {\vspace{12pt}
        \refstepcounter{subappendixc}
        \noindent{\bf Appendix \thesubappendixc. {\kern1pt \bfit #1}}
    \par\vspace{5pt}}
\newcommand{\subsubappendix}[1] {\vspace{12pt}
        \refstepcounter{subsubappendixc}
        \noindent{\rm Appendix \thesubsubappendixc. {\kern1pt \tenit #1}}
    \par\vspace{5pt}}

\newcommand{\textlineskip}{\baselineskip=13pt}
\newcommand{\smalllineskip}{\baselineskip=10pt}

\def\eightcirc{
\begin{picture}(0,0)
\put(4.4,1.8){\circle{6.5}}
\end{picture}}
\def\eightcopyright{\eightcirc\kern2.7pt\hbox{\eightrm c}}

\newcommand{\copyrightheading}[1]
    {\vspace*{-2.5cm}\smalllineskip{\flushleft
    {\footnotesize International Journal of Modern Physics A, #1}\\
    {\footnotesize $\eightcopyright$\, World Scientific Publishing
     Company}\\
     }}

\def\abstracts#1#2#3{{
    \centering{\begin{minipage}{4.5in}\baselineskip=10pt\footnotesize
    \parindent=0pt #1\par
    \parindent=15pt #2\par
    \parindent=15pt #3
    \end{minipage}}\par}}

\renewenvironment{thebibliography}[1]
    {\frenchspacing
     \ninerm\baselineskip=11pt
     \begin{list}{\arabic{enumi}.}
    {\usecounter{enumi}\setlength{\parsep}{0pt}
     \setlength{\leftmargin 12.7pt}{\rightmargin 0pt} 
     \setlength{\itemsep}{0pt} \settowidth
    {\labelwidth}{#1.}\sloppy}}{\end{list}}

\newcounter{itemlistc}
\newcounter{romanlistc}
\newcounter{alphlistc}
\newcounter{arabiclistc}

\newcommand{\fcaption}[1]{
        \refstepcounter{figure}
        \setbox\@tempboxa = \hbox{\footnotesize Fig.~\thefigure. #1}
        \ifdim \wd\@tempboxa > 5in
           {\begin{center}
        \parbox{5in}{\footnotesize\smalllineskip Fig.~\thefigure. #1}
            \end{center}}
        \else
             {\begin{center}
             {\footnotesize Fig.~\thefigure. #1}
              \end{center}}
        \fi}

\newcommand{\tcaption}[1]{
        \refstepcounter{table}
        \setbox\@tempboxa = \hbox{\footnotesize Table~\thetable. #1}
        \ifdim \wd\@tempboxa > 5in
           {\begin{center}
        \parbox{5in}{\footnotesize\smalllineskip Table~\thetable. #1}
            \end{center}}
        \else
             {\begin{center}
             {\footnotesize Table~\thetable. #1}
              \end{center}}
        \fi}

\def\@citex[#1]#2{\if@filesw\immediate\write\@auxout
    {\string\citation{#2}}\fi
\def\@citea{}\@cite{\@for\@citeb:=#2\do
    {\@citea\def\@citea{,}\@ifundefined
    {b@\@citeb}{{\bf ?}\@warning
    {Citation `\@citeb' on page \thepage \space undefined}}
    {\csname b@\@citeb\endcsname}}}{#1}}

\newif\if@cghi
\def\cite{\@cghitrue\@ifnextchar [{\@tempswatrue
    \@citex}{\@tempswafalse\@citex[]}}
\def\citelow{\@cghifalse\@ifnextchar [{\@tempswatrue
    \@citex}{\@tempswafalse\@citex[]}}
\def\@cite#1#2{{$\null^{#1}$\if@tempswa\typeout
    {IJCGA warning: optional citation argument
    ignored: `#2'} \fi}}

\def\pmb#1{\setbox0=\hbox{#1}
    \kern-.025em\copy0\kern-\wd0
    \kern.05em\copy0\kern-\wd0
    \kern-.025em\raise.0433em\box0}

\def\fnt#1#2{\footnotetext{\kern-.3em
    {$^{\mbox{\scriptsize #1}}$}{#2}}}

\def\fpage#1{\begingroup
\voffset=.3in
\thispagestyle{empty}\begin{table}[b]\centerline{\footnotesize #1}
    \end{table}\endgroup}

\def\runninghead#1#2{\pagestyle{myheadings}
\markboth{{\protect\footnotesize\it{\quad #1}}\hfill}
{\hfill{\protect\footnotesize\it{#2\quad}}}}
\font\tenrm=cmr10
\font\tenit=cmti10
\font\tenbf=cmbx10
\font\bfit=cmbxti10 at 10pt
\font\ninerm=cmr9

\font\eightrm=cmr8

\def\qed{\hbox{${\vcenter{\vbox{            
   \hrule height 0.4pt\hbox{\vrule width 0.4pt height 6pt
   \kern5pt\vrule width 0.4pt}\hrule height 0.4pt}}}$}}

\renewcommand{\thefootnote}{\fnsymbol{footnote}}    

\newcommand{\ET}{\mbox{$E_{T}$} }

\newcommand{\EA}{\mbox{$\eta$} }

\newcommand{\AEA}{\mbox{$|\eta|$}}

\def\simge{\mathrel{\rlap{\raise 0.53ex \hbox{$>$}}{\lower 0.53ex \hbox{$\sim$}}}}
\def\simle{\mathrel{\rlap{\raise 0.53ex \hbox{$<$}}{\lower 0.53ex \hbox{$\sim$}}}}
\begin{document}

\runninghead{A QCD Survey: $0 \leq Q^2 \leq 10^5 $ GeV$^2$} {A QCD
Survey: $0 \leq Q^2 \leq 10^5 $ GeV$^2$}

\normalsize\textlineskip
\thispagestyle{empty}
\setcounter{page}{1}

\copyrightheading{}         

\vspace*{0.88truein}

\fpage{1} \centerline{\bf A QCD Survey: $0 \leq Q^2 \leq 10^5 $ GeV$^2$}
\vspace*{0.37truein}
\centerline{\footnotesize Gerald C. Blazey}
\vspace*{0.015truein}
\centerline{\footnotesize\it Physics Department, Northern Illinois
University } \baselineskip=10pt \centerline{\footnotesize\it
DeKalb, Illinois 60115-2854, U.S.A}
\setcounter{footnote}{0}

\vspace*{0.21truein} \abstracts{Selected, recent results,
primarily from collider experiments but including some fixed
target experiments, are presented as a survey of Quantum
Chromodynamics (QCD).  The concepts of leading order and
next-to-leading order QCD are introduced. Inclusive $p\bar{p}$ jet
and dijet production and deep inelastic $ep$ scattering at very
large momentum transfer are shown to be in good agreement with
perturbative QCD (pQCD). Dijet, three-jet and multi-jet results
from $p\bar{p}, ep$, and $ee$ colliders at moderate $Q^2$ are also
compared to pQCD.  BFKL searches from all three colliders are
discussed. Recent measurements of structure functions and
contributions to the parton distribution functions are presented.
New measurements of $\alpha_s$ are summarized, the world average
is $\alpha_s = 0.1184 \pm 0.0031$.}{}{}
\renewcommand{\thefootnote}{\alph{footnote}}

\vspace*{1pt}\textlineskip  
\section{Introduction and a Brief Introduction to QCD}    
\vspace*{-0.5pt}

\noindent Quantum chromodynamics (QCD) beautifully describes the
strong interaction over a surprisingly broad range of phenomena -
from the non-perturbative regime of hadronic structure to the
highest energies observed in particle colliders. Recent
measurements of high momentum transfer jet production and deep
inelastic scattering, multi-jet production, proton parton
distribution functions, and the strong coupling constant
convincingly illustrate the breath of experimental progress. These
results have, in turn, stimulated impressive theoretical progress
and sensitive searches for new interactions and dynamics.  Because
of the immense quantity of QCD research underway the results
discussed here are necessarily and unfortunately incomplete, this
document might best be viewed as an introductory overview.

The proton--antiproton interaction, a general scattering process,
nicely introduces the concepts of leading order and
next--to-leading order perturbative quantum chromodynamics.
Inelastic scattering between a proton and an antiproton can be
described as an elastic collision between a single proton
constituent and single antiproton constituent. These constituents
are collectively referred to as partons and in QCD are quarks and
gluons. Predictions for jet production are given by folding
experimentally determined parton distribution functions $f$ with
quark and gluon two--body scattering cross sections
$\hat{\sigma}$. The two ingredients can be formally combined to
calculate any cross section of interest: $\sigma = \sum_{i,j} \int
dx_{1}dx_{2}
f_{i}(x_{1},\mu^{2}_{F})f_{j}(x_{2},\mu^{2}_{F})\newline \sigma [
x_{1}P, x_{2}P, \alpha_{s}(\mu^{2}_{R}),
Q^{2}/\mu^{2}_{F},Q^{2}/\mu^{2}_{R}]$. The nonperturbative parton
distribution function $f_{k}(x_{l},\mu^{2}_{F})$ describes the
momentum fraction $x$ of the beam momentum $P$ carried by the
$l^{th}$ parton of type $k$. The hard two--body interaction
between the gluon and quark partons can be calculated with
perturbative QCD (pQCD) and is a function of the perturbative or
strong coupling constant $\alpha_{s}$, the hard scale momentum
transfer between incoming particles $Q$, and the renormalization
scale $\mu_{R}$.

The factorized scattering has been illustrated with a leading
order (LO) process; that is, a process with the minimum number of
vertices (or coupling constants) to describe two final states.
Although useful, the leading order picture (where one parton
results in one jet) is too simple and has an unphysical dependence
on $\mu_{R}$. Next--to--leading order (NLO), or for this two jet
process O$(\alpha^{3}_{s})$ calculations, include additional
parton emission and have reduced sensitivity to $\mu_{R}$.
Depending on the proximity of the outgoing partons, a ``jet''
could result from one or the combination of two partons. NLO
calculations crudely model fragmentation, thereby obviating the
need for fragmentation functions.

\section{QCD at the Highest Momentum Transfers}

\begin{figure}[b!] 
\vspace{-1.0in}
\centerline{\epsfig{file=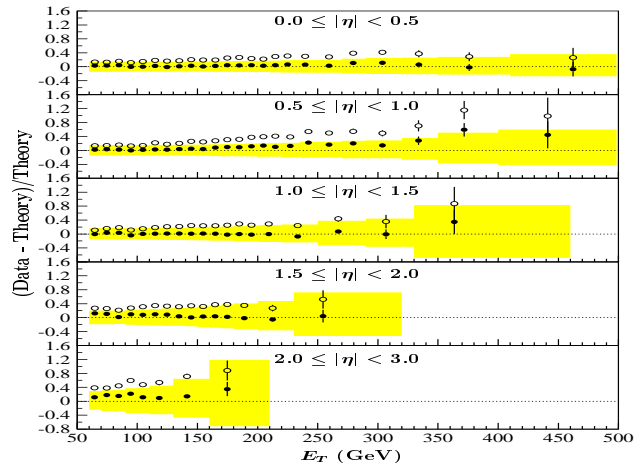,height=4.0in,width=4.75in}}
\vspace{-0.85in}
 \caption{The difference between the inclusive
cross section and NLO theory normalized to NLO theory.}
\label{D0JETSLIN}
\end{figure}

\noindent A complete theoretical description of inclusive jet
production, $p\bar{p} \rightarrow j+X$, requires proper treatment
of the final state radiation and accurate measurements of the
parton distribution functions, pdf's. The inclusive jet cross
section is reported as $d^{2}\eta/dE_{T}d\eta$. $\ET = Esin\theta$
where $E$ is jet energy and $\theta$ the angle between the proton
direction and the jet. The pseudorapidity, $\eta$, is defined as
$-ln(tan(\theta/2))$. Kinematically, an individual jet is
characterized by \ET, $\eta$, and azimuth $\phi$. Jets are found
by clustering energy in a cone of radius 0.7 in $\eta-\phi$
space~\cite{JETALGS}.

The inclusive jet cross section has been intensively studied at
beam energies of 900 GeV/c at the Fermilab Tevatron
proton--antiproton collider in Batavia, Illinois by the
D\O~collaboration for $\AEA < 3.0$~\cite{D0JETS2000,D0JETS1995}
and by the CDF collaboration in the region $0.1 < \AEA <
0.7$~\cite{D0JETS2000,CDFPRL}. The data now span nearly ten orders
of magnitude for jet energies between 20 and 450 GeV. The
percentage difference as a function of \ET between the data and
theory normalized to the theory for the D\O~data is shown in
Fig.~\ref{D0JETSLIN}. Events at the highest \ET correspond to $x$
values of 0.5 and $Q^2$ greater than $10^5$ GeV$^2$. Data points
include statistical errors only and the bands indicate the
magnitude of the systematic errors. The figure includes a NLO
prediction~\cite{GGKEKS} using the CTEQ4HJ pdf. There is excellent
agreement at all \ET and \EA and, in particular, no indication of
excess production at large \ET. The CDF collaboration finds
similar agreement in the region $0.1 < \AEA < 0.7$, but reports a
25\% discrepancy~\cite{D0JETS2000} for \ET above 400 GeV.

The good agreement is in sharp contrast to an earlier, lower
statistics results from the CDF collaboration which suggested
excess production at large \ET relative to contemporaneous NLO
predictions~\cite{CDFPRL}. The early CDF results prompted a closer
look at theoretical and pdf uncertainties contributing to the
predictions. In fact, inspired by the initial CDF result, the
CTEQ4HJ pdf used in Fig.~\ref{D0JETSLIN} includes a strengthened
high-x gluon component. All D\O~and CDF results are consistent
within statistical and systematic errors~\cite{ANNREV}.

\begin{figure}[b!] 
\vspace{-.2in}
\centerline{\epsfig{file=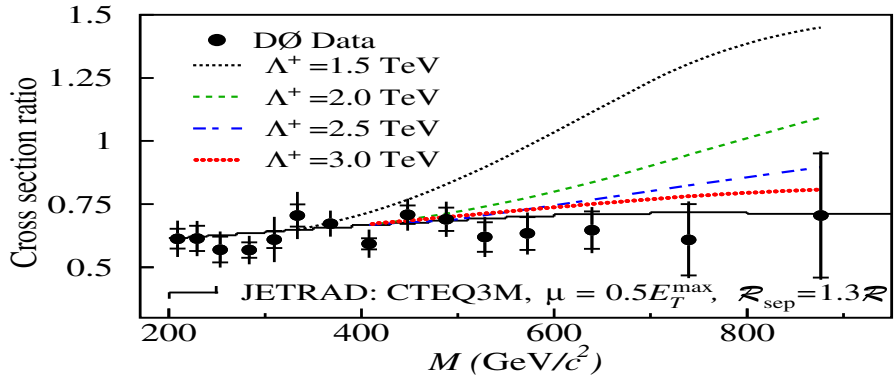,height=2.0in,width=4.75in}}
\caption{Comparisons of central to forward dijet cross section
ratios as a function of mass to theoretical predictions. See text
for details.} \label{D0DIJETLIN}
\end{figure}

The dijet mass distribution for the leading two jets of a
$p\bar{p} \rightarrow j+j+X$ event at beam energies of 900 GeV
also constitutes a sensitive search for new physics. The cross
section is reported as $d^{3}\eta/dM_{jj}d\eta_{1}d\eta_{2}$ where
$M_{jj}$ is the invariant mass of the leading two jets. Both
D\O~and CDF have published results which show good agreement
between data and NLO QCD~\cite{D0ANG95,D0DIJETMASS,CDFDIJETMASS}.
As shown in Fig.~\ref{D0DIJETLIN}, the D0 collaboration has taken
the ratio of the central dijet mass distribution $\AEA \leq 0.5$
to the forward distribution, $0.5 \leq \AEA \leq 1.0$.  This
distribution is sensitive to quark compositeness because the
associated jet production will be predominantly central in
rapidity.  In a manner completely analogous to Rutherford
scattering, excess jet production at very large transverse
energies signals the presence of quark compositeness. The curves
in Fig.~\ref{D0DIJETLIN} include NLO predictions with and without
a additional jet production due to compositeness.  The solid curve
postulates no substructure, and D0 has set a limit of 2.4 TeV or
$\sim 10^{-19}$ m on any substructure~\cite{D0DIJETMASS}.

HERA, an electron-proton collider in Hamburg, Germany provides a
complementary and comprehensive opportunity to examine QCD over a
very large range of momentum transfer (0 to $3\times 10^{4}$
GeV${^2}$). The introductory description of $p\bar{p}$ jet
production can be modified to $ep$ scattering if one incoming
hadron is replaced with an electron and the exchanged gluon by an
electroweak boson . A neutral current (NC) reaction is
characterized by an exchanged photon or $Z$, while charged current
(CC) reactions involve an exchanged $W$. The CC and NC cross
sections are reported as $d^{2}\sigma/dQ^{2}$. HERA ran from
1994--1997 with positrons at a center-of-mass energy of 300 GeV
and from 1997-1998 with electrons at an energy of 320 GeV.

\begin{figure}[b!] 
\vspace{-.375in}
\centerline{\epsfig{file=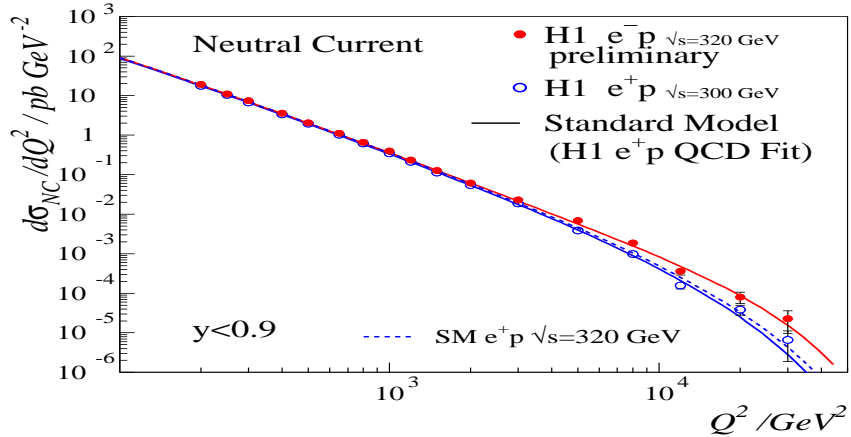,height=2.5in,width=4.75in}}
\vspace{-.125in} \caption{Neutral current cross sections for
$e^{+}p$ and $e^{-}p$ scattering compared to Standard Model
expectations.} \label{H1LOG}
\end{figure}

As shown in Fig.~\ref{H1LOG}, the $ep \rightarrow e+X$ neutral
current cross section spans eight orders of
magnitude~\cite{H1NEW}. Below $Q^2 = 10^{4}~$GeV$^2$ the data are
well described by NLO theory. However, the NC $e^{+}p$ data shows
excess production at very high $Q^2$, an effect observed (and
which generated great excitement) by both the H1 and ZEUS
collaborations in the 1994--1997 data set~\cite{ZEUSHIGH,H1HIGH} .
In contrast, the 1997--1998 $e^{-}p$ NC cross section (also shown
in Fig.~\ref{H1LOG}) and similar results from the ZEUS
collaboration~\cite{ZEUSNEW} are both well described within
statistical errors at all $Q^2$ by NLO QCD. Although not shown, CC
$ep \rightarrow \nu +X$ cross sections for both electrons and
positrons are also well described by QCD~\cite{H1NEW,ZEUSNEW}.

The $p\bar{p}$ and $ep$ colliders both test the limits of QCD and
the Standard Model (SM) at the highest momentum transfer.  In the
mid-nineties hints of excess jet and NC events at large \ET or,
equivalently, $Q^2$ were observed at both colliders.  However,
since then new results (characterized by higher statistics or new
projectiles) and modified theoretical predictions (characterized
by new pdfs), strongly suggest the SM can currently describe the
data. The lure of new physics is exciting and searches for high
$Q^2$ discrepancies will continue, especially with the ever
increasing statistics at HERA and Fermilab. However, as seen in
the next section, the lower $Q^2$ regime of multi-jet production
proves an interesting window onto the behaviour of QCD.

\section{Multijet Production}

\noindent  All three colliders, HERA, the Tevatron, and the LEP
$e^{+}e^{-}$ machine in Geneva, Switzerland prove prolific
laboratories for the study of multi-jet production.  At HERA and
LEP jets are conventionally found with recombination algorithms
which cluster energy according to their proximity in \ET, \EA, and
azimuth~\cite{JETALGS}.  At LO, HERA dijet production occurs
through the QCD Compton graph (where a photon from the incoming
lepton scatters off a quark from the hadron to produce a gluon jet
and quark jet) and through the boson-gluon fusion graph (where a
gluon from the hadron splits into a $q\bar{q}$ pair, one of which
scatters off a photon from the incoming lepton, to produce a quark
and antiquark jet). The $\sim100 < Q^2 < \sim10^4$ GeV, rapidity,
mass, and $x$ distributions of these dijet cross sections are
generally well described by NLO QCD~\cite{HERADIJET}.

\begin{figure}[b!] 
\vspace{-0.25in}
\centerline{\epsfig{file=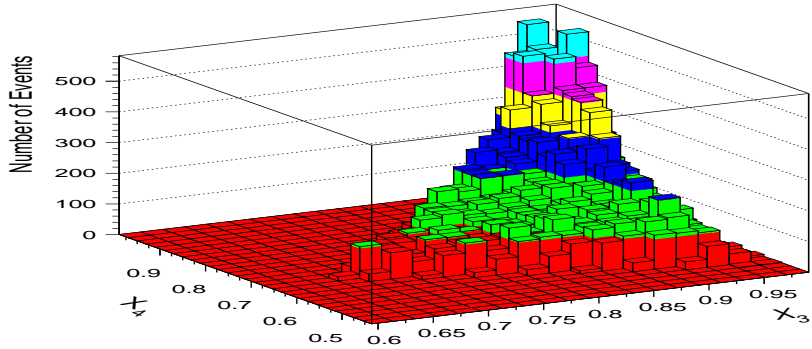,height=2.0in,width=4.75in}}
\vspace{-.125in}
 \caption{A Dalitz plot of a NLO three jet cross
section.} \label{THREEJET}
\end{figure}

The NLO dijet predictions, which include up to three emitted
partons, represent LO predictions for three jet production. These
LO, three jet predictions are in reasonable agreement with HERA
three jet cross sections as a function of $Q^2$ and
$M_{3jet}$~\cite{HERATRIJET}. There does seem to be a discrepancy
between data and theory for the ratio of three to two jet cross
sections. At $Q^2$ below 1000 GeV$^2$ the experimental ratio is
below the LO prediction~\cite{HERATRIJET} . However, parton shower
simulations show excellent agreement with the data, this is
typically a signal that higher order perturbative QCD predictions
will show improved agreement.  The ratio of three jet to two jet
cross sections has also been measured at the Tevatron as a
function of the sum of the \ET of all jets in the event
$\Sigma\ET$~\cite{TEVRATIO}. The ratio rises rapidly with
$\Sigma\ET$ but levels off at 0.7 above 200 GeV.  The data is
fairly well described by LO QCD, but as expected is sensitive to
the choice of renormalization scale.

Recently a theoretical milestone was reached with the completion
of NLO three jet predictions~\cite{THREEJETTHEORY}. At NLO three
jet predictions includes three and four parton final states.
Figure~\ref{THREEJET} shows a Dalitz lego plot calculated with NLO
QCD , where $x_{i} = 2E_{i}/M_{3J}$ and the indices $i=3,4,5$
represent the final state jets ordered in energy, $E_i$ is the
energy of the three jets, and $M_{3J}$ the invariant mass of the
jets.  The pronounced peaking a $x_{4} = x_{3} \sim 1$ indicates
that the third jet is of low energy or that the events are
primarily two jet in nature. The CDF collaboration has made
preliminary measurements of the Dalitz distributions and reports
good agreement, however uncertainty analysis are
pending~\cite{CDFTHREEJET}.

Multijet production also offers an opportunity to search for new
dynamics.  Almost all QCD data are well described by DGLAP
evolution which can be characterized by partonic emission strongly
ordered in $Q^2$. The familiar expansion $\sigma \sim (\alpha_{s}
ln Q^{2})^{n}$ describes a remarkable range of $x$ and $Q^2$. (See
the discussion of Structure Functions, for example.) However, at
fixed $Q^2$ and small x, Balitsky-Fadin-Kuraev-Lipatov or BFKL
evolution may be appropriate~\cite{BFKL}. Here $1/x$ becomes large
and cross sections are of the form $\sigma \sim (\alpha_{s}
ln(1/x))^{n}$ where $x=Q^{2}/s$ and $s$ is the center-of-mass
energy.  Terms in $\alpha_{s} ln(1/x)$ correspond to gluon
emission strongly ordered in $x$.

At the Tevatron, for events characterized by two jets widely
separated in $\Delta\eta$, the term $\alpha_{s} ln(1/x)$ is
proportional to $\alpha_s \Delta \eta$. BFKL
resummation\cite{TEVBFKLTHEORY} of these leading log terms yields
an exponential increase of $\sigma$ with $\Delta \eta$, $\sigma
\sim e^{\Delta \eta}$, this approximation is valid only for
$\Delta\eta
> 2$. Crudely, a gluon ladder between the outgoing partons increases
the cross section. However, the exponential growth is moderated by
momentum conservation (equivalently the pdf's). Sensitivity to the
exponential increase can be found in the ratio of cross sections
at different center-of-mass energies but identical $x_1$ and $x_2$
so that the pdf damping cancels. This ratio has been measured at
Tevatron energies\cite{TEVBFKLEXP} of 1800 and 630 GeV. The data
for $\Delta\eta> 2$ is well above and inconsistent with exact LO
and HERWIG predictions. (Herwig is a LO simulation which includes
fragmentation and hadronization). The BFKL prediction best
describes the data but is also 3$\sigma$ below the observation.

HERA is a natural site for BFKL searches since $x$ values near
$10^{-5}$ are accessible.  For example, inclusive $\pi^{0}$
production for $10^{-4} < x < 10^{-2}$ and $2 < Q^{2} < 70$
GeV$^2$ is well described by LO BFKL predictions but poorly
described by LO QCD~\cite{HERAPION}.  Likewise forward jet
production or production in the beam direction, $1.5 < \eta < 2.8$
and $10^{-4} < x < 10^{-2}$ is not described by NLO
QCD~\cite{HERABFKLJET}.  There is also a pronounced dependency on
the choice of renormalization scale.  This suggests that the
calculations are incomplete and something more is needed, perhaps
BFKL improvements, a better description of the photon structure
function, or next-to-next-to-leading order terms.

BFKL dynamics are also accessible at LEP through
$\gamma^{*}\gamma^{*}$ scattering where each incoming lepton
radiates a virtual photon. At LO the photons scattering off a
$q\bar{q}$ pair.  At higher orders each photon emits a $q\bar{q}$
pair and the two pairs interact through gluon exchange.   The L3
experiment at LEP has measured a rise in
$\sigma(\gamma^{*}\gamma^{*})$ as a function of the scattering
center-of-mass~\cite{LEPBFKL}.  This increased cross section can
be attributed to the the phase space opened up by emission of
additional gluons from the exchanged gluon. The rise is not
described by LO QCD but is described by a BFKL prediction. There
are abundant hints for the existence of BFKL dynamics; however,
all searches need improved higher-order calculations, both DGLAP
and BFKL, before the origins of these hints are clear. Moreover,
DGLAP has been impressively successful over almost all $x$ and
$Q^2$.

\section{The Structure of the Proton}

\noindent The proton pdf's are derived from global fits to two
general types of data: structure functions from $e$, $\mu$ and
$\nu$ scattering from nuclear targets and from exclusive state
production such as Drell--Yan, charm, W, jet, and photon
production.  Without going into details, trial pdf's are evolved
form a starting $Q_o$, convoluted with QCD calculations of the
hard scatter, and fit to the observed data. These pdf's are
assumed to be universal in that a single unique pdf is appropriate
for all reactions. Each reaction does, however, provide access to
a particular type of pdf.  For instance, W production in
$p\bar{p}$ scattering directly accesses the u and d quark pdfs.
Generally, quark functions are heavily constrained by DIS
scattering for $x < 0.9$. Until recently there has been very
little to constrain gluons \cite{HIGHPTGX} above $x > 0.2$ (This
freedom permitted the formulation of pdf's similar to CTEQ4HJ used
in Fig.~\ref{D0JETSLIN}.) Recent developments in pdf derivations
include an increased accuracy due to new types of input data,
replacement of prompt photon data with jet data, and concerted
efforts to improve the treatment of uncertainties.  For a summary
consult the references~\cite{DIS99}.

\begin{figure}[b!] 
\vspace{-.125in}
\centerline{\epsfig{file=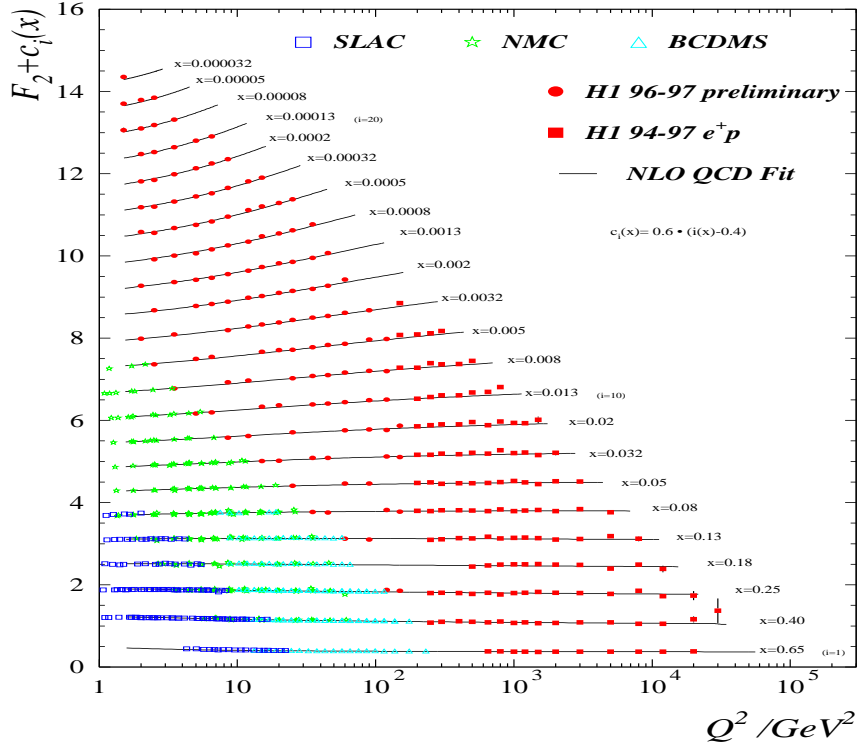,height=4.0in,width=4.75in}}
 \vspace{-.1in}
 \caption{A compilation of structure functions over a wide range of
$x$ and $Q^2$.} \label{SCALING}
\end{figure}

The differential cross section $d^{2}\sigma^{NC}/dxdQ^{2}$ for NC
scattering, where $x$ is the momentum fraction of the struck
quark, can be expressed at leading order in terms of the structure
functions $F_{2}(x,Q^{2})$ and $F_{3}(x,Q^{2})$.  $F_2$ is
proportional to the sum of the valence and sea quark pdf's,
$q(x,Q^{2})+ \bar{q}(x,Q^{2})$. $F_3$ is proportional to the
difference of the valence and sea quark pdf's, $q - \bar{q}$.
Figure~\ref{SCALING} shows $F_2$ for $0.000032 < x < 0.65$ and $1
< Q^{2} < \sim 10^{4}$ GeV$^2$ as measured by the H1, SLAC, NMC
and BCDMS collaborations.  (To aid the eye the data at each value
of x is offset by an arbitrary constant.)  The H1 data is recent,
and ZEUS has made similar measurements~\cite{SCALINGCOMP}. The
curves correspond to a NLO QCD fit and include both $e^{+}p$ and
$e^{-}p$ data . The astonishing agreement includes both the large
Q$^2$ perturbative region and also the small, presumably,
nonperturbative Q$^2$ regions. The dependence of $F_2$ on $Q^2$ at
low-x region is an example of QCD scaling violations.  Although
not shown, the availability of both $e^{+}p$ and $e^{-}p$ NC data
also permits measurement of $F_3$, these results are in agreement
with NLO QCD~\cite{F3}.

\begin{figure}[b!] 
\vspace{-.375in}
\centerline{\epsfig{file=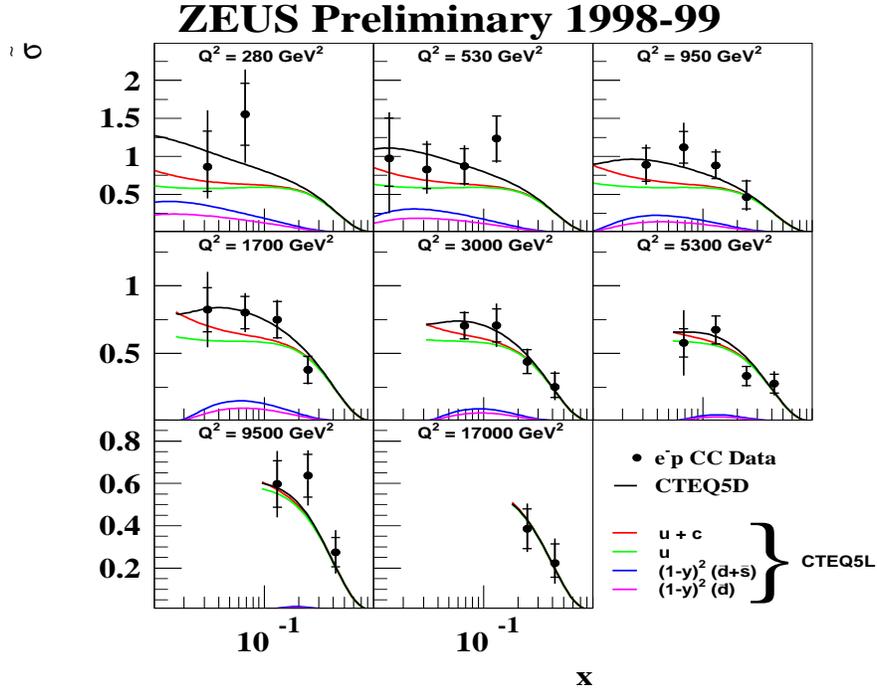,height=4.0in,width=4.75in}}
 \vspace{-.25in}
 \caption{The CC cross-section at various x and $Q^2$ with
predictions for the various quark contributions.}
\label{CCSTFUNC}
\end{figure}

Charge current reactions couple directly to the quark
distributions. Specifically, $d^{2}\sigma^{CC}(e^{-}p)/dxdQ^{2}$
is proportional to the u and c quark distributions, and
$d^{2}\sigma^{CC}(e^{+}p)/dxdQ^{2}$ is proportional to the d
quark.  Figure~\ref{CCSTFUNC} shows a reduced CC cross section for
various $Q^2$ bins as a function of $x$ for $e^{-}p$
data~\cite{HERAQ}. The uppermost curve in each panel represents
the total cross section as predicted by NLO QCD using the CTEQ5D
pdf. Agreement is good at all $x$ and $Q$. The lower curves
indicate the contribution from each type of quark. As expected the
u quark contribution is dominant.

Additional constraints on the quark distributions below $x \sim
0.7$ can be derived from fixed target and $p\bar{p}$ scattering. A
few recent examples are listed here.  The CCFR collaboration has
re-analyzed its $\nu$ and $\bar{\nu}$ DIS data and finds $F_2$ for
$0.015 < x < 0.175$ and $1 < Q^{2} < 100$ GeV$^2$ well described
by updated NLO QCD predictions~\cite{CCFR}. CCFR has also made the
first measurement of $F_{3}^{\nu} - F_{3}^{\bar{\nu}}$.  The
rapidity distributions of the final state leptons in the
Drell--Yan process $p\bar{p} \rightarrow \gamma^{*}/Z \rightarrow
e^{+}e^{-}$  is sensitive to the valence quark distributions. For
example, if the two initial partons are of unequal momentum the
virtual $\gamma^{*}/Z$ will be boosted in the direction of the
more energetic parton and the decay electrons will likewise be
boosted.  Analysis of the rapidity distributions as recently
measured by the CDF collaboration constrains the u and d
distributions for $0.05 < x < 0.6$~\cite{DRELLYAN}. Recent results
also pertain to the sea-quark distributions, the cross sections
for $\mu^{+}\mu^{-}$ production from $pp$ and $pd$ scattering as
measured by the E866 collaboration constrains the $\bar{d} -
\bar{u}$ sea distribution between $x \sim 0.03$ and $\sim
0.3$~\cite{DIS99}.

As mentioned earlier, the large-x gluon distributions are not well
known.  Until recently, a favored process for constraining $g(x)$
in the global fits was direct photon production. The associated
Compton graph samples $g(x)$ directly.  However, this process has
fallen out of favor since NLO calculations cannot describe the
world data~\cite{KT}.  For nearly every photon cross section
measurement, the data exceeds theory at the low end of that
specific measurement's $x$ range. This discrepancy has been
attributed to initial-state soft radiation which manifests itself
as transverse momentum or $k_T$. If not interpreted correctly,
this $k_T$ or ''kick'' adds to the measured $x$ of the photon
which creates the observed excess of data over theory. Evidence
for $k_T$ comes from events where object pairs such as pions,
diphotons, dimuons, and dijets are produced. The vector sum of the
pair momentum is a direct measure of the parton scatter transverse
kick. Indirect evidence comes from the observation that
augmentation of NLO predictions with a phenomenological addition
of $k_T$ dramatically improves agreement between data and theory.
Also, the importance of soft gluon resummation for Drell-Yan,
diphoton, W, and Z production was recognized some time ago.
Similar calculations for direct photon production are underway and
already show much better agreement between data and theory.
Perhaps with time and higher orders photon data will return to the
pdf fits.

\begin{figure}[b!] 
\vspace{-.125in}
\centerline{\epsfig
{file=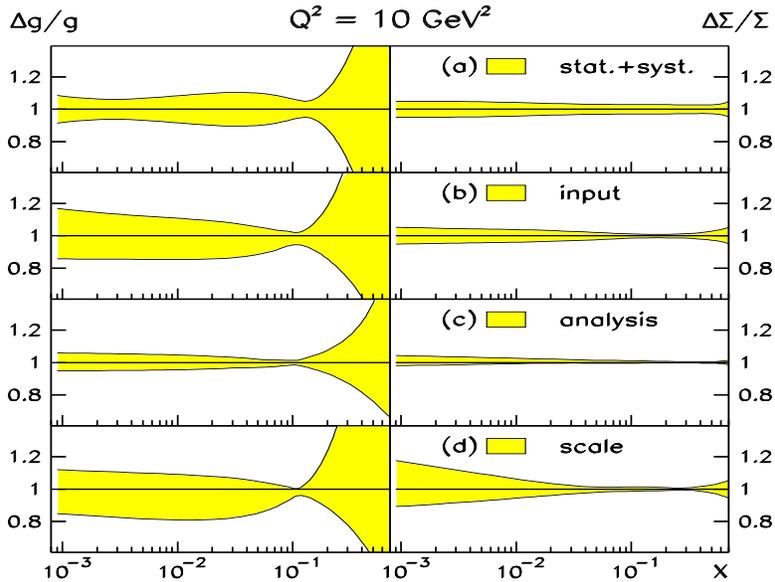,height=3.0in,width=4.0in}}
\caption{Allowed variations of the gluon and quark pdf's assuming
symmetric Gaussian uncertainties associated with the global fit.}
\label{PDFERRORS}
\end{figure}

In the meantime, the central inclusive jet cross sections
constrain $g(x)$ for $\sim0.05 < x < \sim0.5$.  Inclusion of the
full inclusive cross section for $p\bar{p}$ scattering over $\AEA
< 3.0$ will cover $x < \sim 0.8$~\cite{D0JETS2000}.  Even more can
be gained from triple differential cross sections, $d^{3}\sigma /
d\ET d\eta_{1} d\eta_{2}$ where \ET corresponds to the leading jet
in an event and the subscripts identify the leading two
jets~\cite{ANNREV}.  For a dijet event if both jets are at
$\eta=0$ with $\ET = 180$ GeV then $x_1 = x_2 = 0.2$. However, for
$\eta=0$ and $\ET= 90$ GeV with the second jet at $\eta=2$,
$x_1=0.2$ and $x_2 = \sim 0.9$.  Since the event structure is
boosted forward a much greater fraction of the initial hadron
momentum is required. Both Tevatron experiments are encouraged to
publish their preliminary differential cross
sections~\cite{ANNREV,CDFTHREEJET} since NLO predictions exists
and the data could be incorporated into the pdf's.  Dijet
production at HERA, through the boson fusion graph, also directly
measures $g(x) < 0.1$~\cite{HERAGX}. Although the data doesn't
greatly alter the pdfs's relative to the $F_2$ data it does add
stability to the fits and reduce errors.

Until recently the uncertainties of observables due to pdf's were
estimated merely by using a menu of pdf's.  However, pdf's should
more properly handle statistical and systematic data
uncertainties, theoretical uncertainties, and parameterizations.
In the past year or two, there has been very great progress
dealing with these uncertainties~\cite{BOTJE,GIELE,QCDII}. For
instance, the plot below by Botje shows the allowed variations in
the gluon and quark distributions due to statistical
uncertainties, input choices, analysis cuts, and renormalization
choices assuming all uncertainties are symmetric and Gaussian in
nature~\cite{BOTJE}. Notice the unconstrained nature of the gluon
distributions, Tevatron jet results were not included in this
calculation. An even more general approach by Giele and Keller
incorporates probability distributions to derive the
pdf's~\cite{GIELE}. These authors and others are providing tools
for incorporating all sources of pdf uncertainties.

\section{Status of $\alpha_s$ Measurements}

\noindent An enormous body of research has been dedicated to the
study of the strong coupling constant, $\alpha_{s}$ since it is
the only free parameter of QCD and must be determined
experimentally.  Of equal interest is the dependence of the
coupling constant on momentum transfer $Q^{2}$, $\alpha_{s}(Q^{2})
= 12\pi/(33-n_{f})log(Q^{2}/\Lambda^{2})$ where $n_{f}$ is the
number of quark flavors and $\Lambda$ is experimentally
determined. Notice $\alpha_{s}$ decreases or ``runs'' with
momentum transfer. This is, in fact, the basis for the
perturbative NLO QCD calculations described earlier.

The strong coupling constant can be derived in a myriad of ways
and at all $Q^2$, from absolute decay rates of the $Z$ boson and
$\tau$ lepton, energy levels of bound heavy quarks, jet event
shapes, jet production rates and angular distributions, and
scaling violations in deep inelastic scattering. Notable new
measurements of $\alpha_s$ originate from all three high energy
colliders.  In particular, the LEP collaborations recently ran at
center--of--mass energies from 192 to 202 GeV and from event shape
and jet rate measurements determined an average
$\alpha_{s}(198GeV)= 0.109 \pm 0.001 \pm 0.005$ where the first
error is statistical and the second theoretical \cite{LEP1999}. A
uniform re-analysis of $e^{+}e^{-}$ event shapes, jet rates, and
multiplicities from the JADE and LEP collaborations from
center-of-mass energies between 20 and 200 GeV show the strong
coupling constant obviously running with a value of
$\alpha_{s}(M_{Z})= 0.1208 \pm 0.0006 \pm 0.0048$~\cite{LEP1999}.
Finally, comparison of NLO inclusive jet and dijet cross section
predictions to experimental results from HERA and the Tevatron
consistently yield $\alpha_s(M_{Z})$ measurements near
0.12~\cite{HERAGX,CDFALPHA}.  The Tevatron inclusive jet
measurement beautifully demonstrates the running of $\alpha_s$
over $10^3 < Q^{2} < 2\times10^5$ GeV$^2$.

An excellent review by Bethke includes a comprehensive compilation
of the many derivations of $\alpha_{s}$ complete with a
description and evaluation of each measurement \cite{BETHKE}. The
current world average for the coupling constant,
$\alpha_{s}(M_{Z}) = 0.1184 \pm 0.0031$, is based on six
measurements for which next-to-next-to-leading order predictions
exist. The average is known to better than 3\% and is unchanged
since 1997~\cite{SCHMELLING}.

\section{Conclusions}
\noindent QCD, and in particular perturbative QCD, describes the
strong interaction over an impressive kinematic range. At the
highest $Q^2$ accessible, recent data and analyses reveal few, if
any, excursions from the Standard Model.  At intermediate $Q^2$,
jet and multi-jet studies have permitted a closer study of higher
order process and searches for BFKL dynamics.  These searches are
ambiguous. The proton structure functions are under intense study.
Although the quark components are well constrained, the gluon
component is not well measured for $x>0.2$. Recent jet and, if
rehabilitated, photon production data should improve this
situation. There has also been great progress on pdf error
analysis.  Data from all values of $Q^2$ have contributed to a
precise 3\% measurement of the strong coupling constant.

\nonumsection{Acknowledgements} \noindent Many individuals from
the Tevatron, HERA, LEP and fixed target experiments have assisted
with the preparation of this document.  Space does not permit
individual recognition, many thanks to all. I would also like to
acknowledge the support of the NSF.

\nonumsection{References}

\end{document}